\documentclass[aps,twocolumn,showpacs,byrevtex,prl,reprint,nofootinbib]{revtex4-1}
\usepackage{xspace}
\usepackage{epsfig}
\usepackage{dcolumn}
\usepackage{bm}
\usepackage{color}
\usepackage{overpic}
\usepackage{graphicx}
\usepackage{subfigure}
\usepackage{ulem}
\usepackage{pstricks}
\usepackage{pst-node}
\usepackage{rotating}
\usepackage{times}
\usepackage[english]{babel}
\usepackage{epstopdf}
\addto{\captionsenglish}{

}
\usepackage[colorlinks,linkcolor=blue,citecolor=blue]{hyperref}
\usepackage{lineno}

\newcommand{\bfg}{\begin{figure}}
\newcommand{\efg}{\end{figure}}
\newcommand{\bitm}{\begin{itemize}}
\newcommand{\eitm}{\end{itemize}}
\newcommand{\bnum}{\begin{enumerate}}
\newcommand{\enum}{\end{enumerate}}
\newcommand{\btbl}{\begin{table*}}
\newcommand{\etbl}{\end{table*}}
\newcommand{\btbu}{\begin{tabular}}
\newcommand{\etbu}{\end{tabular}}
\newcommand{\bcl}{\begin{center}}
\newcommand{\ecl}{\end{center}}
\newcommand{\bbt}{\bibitem}

\begin{document}
\normalsize
\parskip=5pt plus 1pt minus 1pt
\title{\boldmath Measurement of cross-section for $e^+e^-\to\Xi^-\bar{\Xi}^+$ near threshold at BESIII}
\author{
M.~Ablikim$^{1}$, M.~N.~Achasov$^{10,c}$, P.~Adlarson$^{67}$, S. ~Ahmed$^{15}$, M.~Albrecht$^{4}$, R.~Aliberti$^{28}$, A.~Amoroso$^{66A,66C}$, Q.~An$^{63,49}$, X.~H.~Bai$^{57}$, Y.~Bai$^{48}$
% ...
\\
\vspace{0.2cm}
(BESIII Collaboration)\\
\vspace{0.2cm} {\it
$^{1}$ Institute of High Energy Physics, Beijing 100049, People's Republic of China\\
% ...
$^{4}$ Bochum Ruhr-University, D-44780 Bochum, Germany\\
% ...
$^{10}$ G.I. Budker Institute of Nuclear Physics SB RAS (BINP), Novosibirsk 630090, Russia\\
% ...
$^{15}$ Helmholtz Institute Mainz, Johann-Joachim-Becher-Weg 45, D-55099 Mainz, Germany\\
% ...
$^{28}$ Johannes Gutenberg University of Mainz, Johann-Joachim-Becher-Weg 45, D-55099 Mainz, Germany\\
% ...
$^{48}$ Southeast University, Nanjing 211100, People's Republic of China\\
$^{49}$ State Key Laboratory of Particle Detection and Electronics, Beijing 100049, Hefei 230026, People's Republic of China\\
% ...
$^{57}$ University of Jinan, Jinan 250022, People's Republic of China\\
% ...
$^{63}$ University of Science and Technology of China, Hefei 230026, People's Republic of China\\
% ...
$^{66}$ University of Turin and INFN, (A)University of Turin, I-10125, Turin, Italy; (B)University of Eastern Piedmont, I-15121, Alessandria, Italy; (C)INFN, I-10125, Turin, Italy\\
$^{67}$ Uppsala University, Box 516, SE-75120 Uppsala, Sweden\\
% ...
$^{c}$ Also at the Novosibirsk State University, Novosibirsk, 630090, Russia\\
% ...
}
}


\begin{abstract}
The Born cross-sections and effective form factors for process $e^+e^-\to\Xi^-\bar{\Xi}^+$ are measured at eight center-of-mass energies between 2.644 and 3.080\,GeV, using a total integrated luminosity of 363.9\,pb$^{-1}$ $e^+e^-$ collision data collected with the BESIII detector at BEPCII. After performing a fit to the Born cross-section of $e^+e^-\to\Xi^-\bar{\Xi}^+$, no significant threshold effect is observed.
\end{abstract}
\maketitle

Electromagnetic form factors, which parameterize the inner structure of hadrons, are important observables for improving our understanding of Quantum Chromodynamics (QCD). In the 1960s, Cabibbo and Gatto first proposed that the time-like electromagnetic form factors can be studied in electron-positron collisions by measuring hadron pair-production cross-sections~\cite{emffs}. Assuming that spin-1/2 baryon pair ($B\bar{B}$) production is dominated by one-photon exchange, the Born cross-section for the process $e^+e^-\to B\bar{B}$ can be parameterized in terms of an electromagnetic, $G_E$, and a magnetic, $G_M$, form factor~\cite{cabibbo} as
\begin{equation}\label{func:bcs}
\sigma^B(s) = \frac{4\pi\alpha^2\beta\mathcal{C}}{3s}\left[\left|G_M(s)\right|^2 + \frac{2m_B^2c^4}{s}\left|G_E(s)\right|^2\right],
\end{equation}
where $\alpha$ is the fine-structure constant, $c$ is the speed of light, $s$ is the square of the center-of-mass (c.m.) energy, $\beta = \sqrt{1-4m_B^2c^4/s}$ is a phase-space factor, and $m_B$ is the mass of the baryon. The Coulomb factor $\mathcal{C}$, which accounts for the electromagnetic interaction of point-like fermions in the final state~\cite{fc1}, is unity for neutral baryon,  and equal to $y/(1-e^{-y})$ for charged baryons, where $y=\pi\alpha\sqrt{1-\beta^2}/\beta$. In Ref.~\cite{sss1}, $\mathcal{C}$ is parameterized as an enhancement factor $\mathcal{E}$ times a resummation factor $\mathcal{R}$, $i.e.$ the so called Sommerfeld-Schwinger-Sakharv rescattering formula: $\mathcal{C = E \times R}$ where $\mathcal{E} = \pi\alpha/\beta$. Thus in the limit of $\beta\to0$, the Coulomb factor tends to $\mathcal{E}$, and the factor of $\beta$ due to phase space is canceled which results in a nonzero cross-section at threshold. The effective form factor, which is defined as a linear combination of the electromagnetic form factors,
\begin{equation}\label{func:eff}
|G_{eff}(s)|=\sqrt{\frac{|G_M(s)|^2+\frac{2m_B^2c^4}{s}|G_E(s)|^2}{1+\frac{2m_B^2c^4}{s}}},
\end{equation}
and, through substitution of Eq.(\ref{func:bcs}) into Eq.(\ref{func:eff}), 
is proportional to the square root of the Born cross-section:
\begin{equation}\label{func:eff1}
|G_{eff(s)}|=\sqrt{\frac{\sigma^B(s)}{\frac{4\pi\alpha^2\beta\mathcal{C}}{3s}(1+\frac{2m_B^2c^4}{s})}}.
\end{equation}

There have been many experimental studies on the nucleon pair production cross-sections over several decades, and unusual behavior has been observed in the near-threshold region. The measured cross-section for $e^+e^-\to p\bar{p}$ is approximately constant in the energy ranging from threshold to about 2\,GeV~\cite{ppbar1, ppbar2, ppbar3}, with an average value of about 0.85\,nb. Similar behavior in the near threshold region was also observed in the $e^+e^-\to n\bar{n}$ process~\cite{nnbar}, with an average cross-section of about 0.8\,nb. The non-vanishing cross-section near threshold and the wide-range plateau have attracted great interest and driven many theoretical studies, including scenarios of invoking $B\bar{B}$ bound states or unobserved meson resonances~\cite{theory1}, Coulomb final-state interactions or the quark electromagnetic interaction and the asymmetry between attractive and repulsive Coulomb factors~\cite{theory2}. In the present context of QCD and our understanding of the quark-gluon structure of hadrons, it is particularly interesting to explore similar anomalous phenomenon in the hyperon system~\cite{exp1, exp2, lam, lamc, whitepaper}. Recently, the BESIII collaboration has measured the Born cross-sections for the processes $e^+e^-\to\Lambda\bar{\Lambda}$~\cite{lam} and $e^+e^-\to\Lambda_c^+\bar{\Lambda}_c^-$~\cite{lamc} with the energy scan technique. The unprecedented precision of these measurements  allowed for an observation of a threshold enhancement effect, $i.e.$ non-vanishing cross-section at threshold, to be observed for the first time. The cross-sections of $\Xi$ baryon pair production have been measured at a series of charmonium resonances~\cite{cleoxi} and above the open charm threshold~\cite{singletag}, but the threshold effect of $\Xi$ pair production has never been investigated before due to the limited sample sizes available.

In this paper, we present a measurement of Born cross-sections and effective form factors for the process $e^+e^-\to\Xi^-\bar{\Xi}^+$ at c.m. energies between 2.644 and 3.080\,GeV and perform a fit to the measured Born cross-sections under various hypotheses. The data set used in this analysis corresponds to a total of 363.9\,pb$^{-1}$ $e^+e^-$ collision data~\cite{dataset} collected with the BESIII detector~\cite{besiii} at BEPCII~\cite{bepcii}.

The BESIII detector is a magnetic spectrometer~\cite{besiii}. The cylindrical core of the BESIII detector consists of a helium-based multilayer drift chamber (MDC), a plastic scintillator time-of-flight system (TOF), and a CsI(Tl) electromagnetic calorimeter (EMC), which are all enclosed in a superconducting solenoidal magnet providing a 1.0\,T magnetic field. The solenoid is supported by an octagonal flux-return yoke with resistive plate counter muon-identifier modules interleaved with steel. The acceptance of charged particles and photons is 93\% over the $4\pi$ solid angle. The charged-particle momentum resolution at 1\,GeV/c is 0.5\%, and the d$E$/d$x$ resolution is 6\% for electrons from Bhabha scattering. The EMC measures photon energies with a resolution of 2.5\% (5\%) at 1\,GeV in the barrel (end-cap) region. The time resolution of the TOF barrel part is 68 ps, while that of the end cap part is 110\,ps. The end-cap TOF system was upgraded in 2015 with multi-gap resistive plate chamber technology, providing a time resolution of 60\,ps.

To determine the detection efficiency, 100,000 $e^+e^-\to\Xi^-\bar{\Xi}^+$ Monte Carlo (MC) events are generated for each energy point by using the \textsc{conexc} generator~\cite{conexc}, which takes into account the beam-energy spread and the correction of initial-state radiation (ISR). The decay of the $\Xi^-$ baryon and its anti-baryon are simulated inclusively via \textsc{evtgen}~\cite{evtgen}. The response of the BESIII detector is modeled with a \textsc{geant}{\footnotesize 4}-based~\cite{geant4} MC package. Generic hadronic events from $e^+e^-$ collision are generated with the \textsc{conexc} generator~\cite{conexc} for background studies~\cite{bkgstudy}. The generic hadronic MC is generated according to the $e^+e^-\to$ hadrons cross-section, and the subsequent decays are processed via \textsc{evtgen}~\cite{evtgen} according to the measured branching fractions provided by the Particle Data Group (PDG)~\cite{pdg}.

As the full reconstruction method suffers from low selection efficiency for the process $e^+e^-\to\Xi^-\bar{\Xi}^+$, a single baryon-tag technique~\cite{cleoxi, singletag} is applied in this analysis. We fully reconstruct the $\Xi^-$ via its $\Lambda\pi^-$ decay mode with $\Lambda\to p\pi^-$, and infer the presence of the anti-baryon $\bar{\Xi}^+$  from the distribution of the  recoiling mass against the reconstructed system (unless otherwise noted, the charge-conjugate state of the $\Xi^-$ mode is included implicitly throughout the paper).

Charged tracks are required to be reconstructed in the MDC with good helical fits and within the angular coverage of the MDC: $|\cos\theta|<0.93$, where $\theta$ is the polar angle with respect to the $e^{+}$ beam direction. Information from the specific energy deposition measured in the MDC, combined with the information on flight time measured in the TOF, are used to form particle identification (PID) confidence levels for the $\pi$/$K$/$p$ hypotheses. Each track is assigned the particle type with the highest confidence level. Events with at least two negatively charged pions and one proton are kept for further analysis. 

To reconstruct $\Lambda$ candidates, a secondary vertex fit~\cite{vtxfit} is applied to all $p\pi^-$ combinations; the ones characterized by $\chi^{2}<500$ are kept for further analysis. The mass resolution of the $p\pi^-$ pair is 1 MeV/$c^{2}$, and the invariant mass of the $p\pi^-$ pair is required to be within 5 MeV/$c^{2}$ of the nominal $\Lambda$ mass from PDG~\cite{pdg}, determined by optimizing the figure of merit ${N_S/\sqrt{N_S + N_B}}$ based on the MC simulation, where $N_S$ is the number of signal MC events and ${N_B}$ is the number of inclusive background events. $N_S$ and $N_B$ are normalized according to the number of signal and background events in data. To suppress background from non-$\Lambda$ events further, the decay length of the $\Lambda$ candidate, i.e. the distance between its production and decay positions, is required to be greater than zero. The $\Xi^{-}$ candidates are reconstructed with a similar strategy using a secondary vertex fit, and the candidate with the minimum value of $|M_{\Lambda\pi^-}-m_{\Xi^{-}}|$ among all $\Lambda\pi^{-}$ combinations is selected, where $M_{\Lambda\pi^{-}}$ is the invariant mass of the $\Lambda\pi^-$ pair, and $m_{\Xi^{-}}$ is the nominal $\Xi^{-}$ mass from the PDG~\cite{pdg}. The invariant mass of the $\Lambda\pi^-$  pair is required to be within 10\,MeV/$c^{2}$ of the nominal $\Xi^{-}$ mass, and the decay length of the $\Xi^{-}$ candidate is required to be greater than zero.

\begin{figure}[!htbp]
	\includegraphics[width=0.45\textwidth]{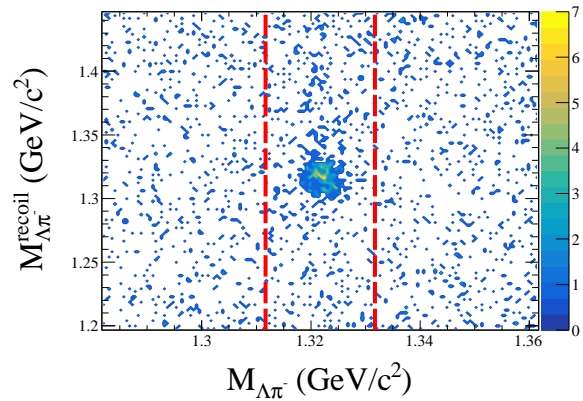}
	\caption{ Distribution of $M^{\rm recoil}_{\Lambda\pi^-}$ versus $M_{\Lambda\pi^-}$ from data. The dashed lines denote the $\Xi^{-}$ signal region.}
	\label{scatterplot}
\end{figure}

The anti-baryon $\bar\Xi^{+}$ candidate  can be inferred from the system recoiling against the selected $\Lambda\pi^-$ pair,
\begin{equation}
M^{\rm recoil}_{\Lambda\pi^-} = \sqrt{(E_{e^+e^-}-E_{\Lambda\pi^-})^{2} - |\vec{p}_{e^+e^-}-\vec{p}_{\Lambda\pi^-}|^{2}c^2},
\end{equation}
where $E_{\Lambda\pi^-}$ and $\vec{p}_{\Lambda\pi^-}$ are the energy and momentum of the $\Lambda\pi^-$ system, and $E_{e^+e^-}$ and $\vec{p}_{e^+e^-}$ are the energy and momentum of the $e^+e^-$ system. Figure~\ref{scatterplot} shows the distribution of $M_{\Lambda\pi^{-}}$ versus $M^{\rm recoil}_{\Lambda\pi^{-}}$ for the data integrated over all eight energy points. The dashed lines denote the $\Xi^-$ signal region. A clear accumulation around the nominal value of the $\Xi$ mass can be seen. Potential sources of background are investigated by studying the generic hadronic MC samples after imposing the signal-selection criteria. It is found that $e^+e^-\to\Sigma^0\pi^-\bar{\Sigma}^+$ is the dominant background process, and is distributed smoothly throughout the region of interest.

\begin{figure}[!htbp]
	\includegraphics[width=0.4\textwidth]{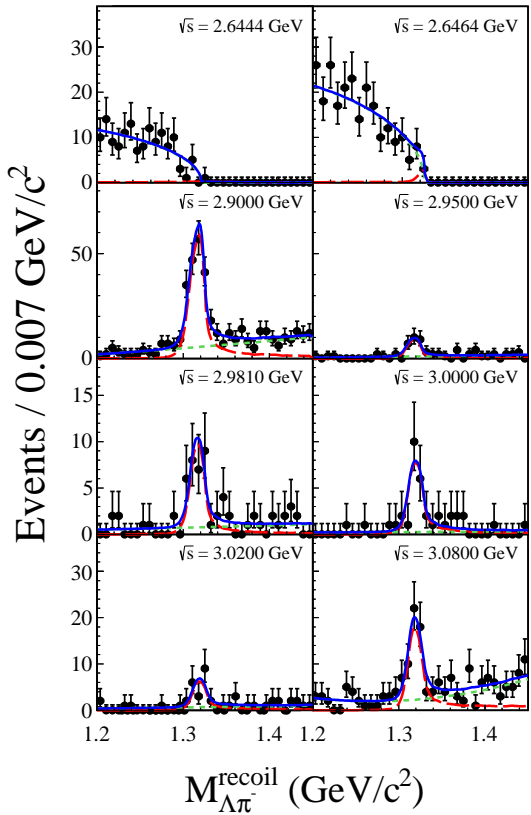}
	\caption{Fit of the recoil mass spectra of $\Lambda\pi^-$ at each energy point. Dots with error bars are data, the solid lines show the fit result, the long dashed lines represent the signal contribution, and the short dashed lines represent the smooth background.}
	\label{fitting}
\end{figure}

The signal yields for $e^+e^-\to\Xi^-\bar{\Xi}^+$ at each energy point are determined by performing an unbinned maximum likelihood fit to the $M^{\rm recoil}_{\Lambda\pi^-}$ spectrum in the range between 1.2\,GeV/$c^{2}$ and 1.45\,GeV/$c^{2}$. The signal shape for the decay $e^+e^-\to\Xi^-\bar{\Xi}^+$ at each energy point is represented by the individual MC-simulated shape convoluted with a Gaussian function. The background is described by a second-order polynomial while for the two energy points near threshold, 2.644 and 2.646\,GeV, the background is described by an Argus function~\cite{argus}. All parameters of the probability density functions are floated in the fit. Figure~\ref{fitting} shows the $M^{\rm recoil}_{\Lambda\pi^-}$ distributions and the fit results for the $e^+e^-\to\Xi^-\bar{\Xi}^+$ at each energy point.

The Born cross-section for $e^+e^-\to\Xi^-\bar{\Xi}^+$ is calculated as
\begin{equation}
\sigma^{B} =\frac{N_{\rm obs}}{{\cal{L}}\cdot\epsilon\cdot(1 + \delta)},
\end{equation}
where $N_{\rm obs}$ is the number of the observed signal events, ${\cal{L}}$ is the integrated luminosity, $\epsilon$ is the detection efficiency of the single baryon-tag technique, including the charge-conjugate state as well as the branching fractions for the subsequent decays of the $\Lambda$ and $\Xi$ particles, and $(1 + \delta)$ is the ISR correction factor~\cite{fisr}, including the vacuum-polarization correction~\cite{fvp}. The product of the ISR correction factor and the detection efficiency, $\epsilon(1+\delta)$,  is obtained via an iteration method~\cite{iteration}, $i.e.$ feeding the measured Born cross-section (Eq.(\ref{func1}) is used to describe the Born cross-section lineshape during the iteration) back into the MC simulation until the result converges at the 1.0\% level. The effective form factor for the process $e^+e^-\to\Xi^-\bar{\Xi}^+$ is calculated with Eq.(\ref{func:eff1}). Table~\ref{sum} summarizes the measured Born cross-sections and effective form factors. For the two energy points near threshold, no clear excess of the signal component is observed and the corresponding upper limits are calculated at 90\% confidence level (C.L.) based on the profile likelihood method~\cite{uplimit}. Here the systematic uncertainties in the efficiency are taken into account in the upper limit calculation. The single-baryon tag method leads to the double counting effect of the $\Xi^-\bar{\Xi}^+$ final state,  which is taken into account when calculating the statistical uncertainties based on the study of MC simulation~\cite{doublecounting}. In this analysis, the double-counting ratio is about 19\%. The double counting does not affect the central value of the final result but does affect the statistical uncertainty. If the uncertainty is determined by fitting, the relative statistical uncertainty is underestimated by about 8\%.  

\btbl[!htbp]
	\caption{\label{sum} 
Summary of measured Born cross-sections $\sigma^{B}$ and effective form factors $|G_{eff}|$ for $e^+e^-\to\Xi^-\bar{\Xi}^+$, where $\sqrt{s}$ is the $e^+e^-$ c.m. energy, $\mathcal{L}$ is the integrated luminosity~\cite{dataset}, $\epsilon(1+\delta)$ is the  product of the detection efficiency and the ISR correction factor, $N_{\rm obs}$ is the number of signal events and $S$ is the signal significance. The values between brackets represent the corresponding upper limit at the 90\% confidence level. The first uncertainty is statistical and the second is systematic. }
	\begin{tabular}{ccccccc}
		\hline\hline
		$\sqrt{s}$ (GeV) &$\mathcal{L}$ (pb$^{-1}$) &$\epsilon(1+\delta)$ & $N_{\rm obs}$ & \hspace*{1.2cm}$\sigma^B$ (pb)\hspace*{1.2cm}  &$|G_{eff}| (\times 10^{-2})$ &$S$($\sigma$)\\ \hline
		2.644  &33.7  &0.015 &2.2$^{ + 2.9}_{ - 1.5}$ ($<$8.5)  &4.4$^{ + 6.2}_{ - 3.2}$ $\pm$ 0.4 ($<$16.8) &7.6$^{ + 5.4}_{ - 2.8}$ $\pm$ 0.4 ($<$15.0) &0.5\\
		2.646  &34.0  &0.022 &4.8$^{ + 4.1}_{ - 2.8}$ ($<$12.8)  &6.4$^{ + 5.9}_{ - 4.0}$ $\pm$ 0.6 ($<$17.1) &7.6$^{ + 3.5}_{ - 2.4}$ $\pm$ 0.4 ($<$12.4) &0.5\\
		2.900  &105  &0.292 &213.5$\pm$21.5 &7.0$\pm$0.8$\pm$0.5   &3.4$\pm$0.2$\pm$0.1 &17\\
		2.950  &15.9  &0.287 &30.5 $\pm$6.3  &6.7$\pm$1.5$\pm$0.4   &3.2$\pm$0.4$\pm$0.1 &6\\
		2.981  &16.1  &0.304 &34.0 $\pm$5.9  &6.9$\pm$1.3$\pm$0.5   &3.3$\pm$0.3$\pm$0.1 &7\\
		3.000  &15.9  &0.323 &28.3 $\pm$5.6  &5.5$\pm$1.2$\pm$0.4   &2.9$\pm$0.3$\pm$0.1 &7\\
		3.020  &17.3  &0.335 &23.1 $\pm$6.5  &4.0$\pm$1.2$\pm$0.3   &2.5$\pm$0.4$\pm$0.1 &5\\
		3.080  &126  &0.323 &70.5 $\pm$11.3 &1.7$\pm$0.3$\pm$0.1   &1.6$\pm$0.1$\pm$0.1 &8\\
		\hline\hline
	\end{tabular}
\etbl

Several sources of systematic uncertainties are considered in the Born cross-section measurement. They are related to the luminosity measurement, the $\Xi$ reconstruction efficiency, the fit procedure, the angular distribution and the ISR-correction factor. The integrated luminosity is measured with a 1.0\% precision~\cite{dataset}. The systematic uncertainty for the $\Xi$ reconstruction efficiency, including the tracking and PID efficiencies, as well as the requirements on the mass window and decay length of the $\Xi/\Lambda$, are studied using the same method as described in Refs.~\cite{singletag, xirec}. For the two energy points near threshold, the uncertainty of $\Xi$ reconstruction efficiency is studied with $J/\psi\to\Xi^-\bar{\Xi}^+$ events, and the momenta of pions in the final state are required to be within the same range as at threshold. The systematic uncertainty due to the fit of the $M^{\rm recoil}_{\Lambda\pi^-}$ spectrum includes considerations of the fitting range, signal shape and background shape. The systematic uncertainty associated with the fitting range is estimated by varying the mass range in steps of 50\,MeV/$c^{2}$. The systematic uncertainty associated with the signal shape is estimated by changing the nominal signal shape to a Gaussian and its parameters are fixed according to the fit of signal MC shape. For the uncertainty due to the background shape, since the background distributes smoothly in the region of interest, it is estimated by performing an alternative fit with a third-order polynomial function. The systematic uncertainty associated with the angular distribution is studied with the same method as described in Ref.~\cite{singletag}, by weighting the cos$\theta_\Xi$ difference for each bin between data and the phase-space MC sample, where $\theta_\Xi$ is the angle between $\Xi$ and the beam direction in the $e^+e^-$ c.m. system. The uncertainty due to the iterative MC tuning procedure is assigned from the difference between the final two iterations. Since the ISR-correction factor is calculated with the measured cross-section, the correlations between energy points are also taken into account by using the method described in Ref.~\cite{isrcorr}. The various systematic uncertainties on the cross-section measurements are summarized in Table~\ref{bcssys}, where the values between brackets represent the corresponding sources of systematic uncertainty for the two energy points near threshold. Assuming all sources to be independent, the total systematic uncertainty is obtained by summing over the individual contributions in quadrature.
\begin{table}[!htp]
	\centering
	{\caption{Summary of sources of the systematic uncertainty on the Born cross-section measurement (in \%). The values in parentheses represent the systematic uncertainty near threshold.
		}
		\label{bcssys}}
	\begin{tabular}{cc}
		\hline\hline
		Source                             &Value\\ \hline
		Luminosity                         &1.0\\
		$\Xi$ reconstruction               &3.4 (7.4)\\
		Fit range                          &3.8\\
		Signal shape                       &0.5\\
		Background shape                   &0.6\\
		Angular distribution               &3.6\\
		ISR  factor                        &1.3\\  \hline
		Total                              &6.5 (9.3)\\  \hline \hline
	\end{tabular}
\end{table}

\begin{figure}[!htbp]
	\includegraphics[width=0.45\textwidth]{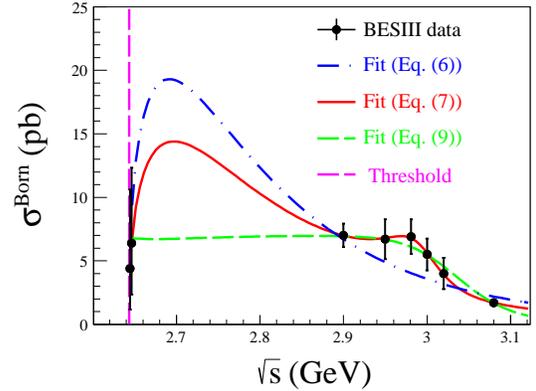}
	\caption{Fit to the measured Born cross-section with different assumptions. The dots with the error bars are the measured Born cross-sections at c.m. energies between 2.644 and 3.080\,GeV. The dash-doted line denotes the fit results using Eq. (\ref{func1}). The solid line denotes the fit results using Eq. (\ref{func11}). The dashed line denotes the fit results using Eq. (\ref{func2}), The vertical dashed line denotes the production threshold for $e^+e^-\to\Xi^-\bar{\Xi}^+$.}
	\label{bcsfit}
\end{figure}

A search for a threshold effect is made by performing a least-$\chi^{2}$ fit to the Born cross-section of $e^+e^-\to\Xi^-\bar{\Xi}^+$ with a series of alternative assumed functions. The first of these is a perturbative QCD-driven energy power function~\cite{fpqcd} which was successfully applied in the $e^+e^-\to\Lambda\bar{\Lambda}$~\cite{lam} and $e^+e^-\to\Sigma^{\pm}\bar{\Sigma}^{\mp}$~\cite{sig} processes, 
\begin{equation}
\sigma^{B}(\sqrt{s}) =\frac{c_0\cdot\beta\cdot\cal{C}}{(\sqrt{s}-c_1)^{10}},
\label{func1}
\end{equation}
where $c_0$ and $c_1$ are free parameters. The blue dash-doted line in Fig.~\ref{bcsfit} shows the fit result with a  $\chi^2$ divided by the number of degrees of freedom ($\chi^2/d.o.f$) equal to 11.87/6. It is seen that this fit does not describe the data points between 2.981 and 3.020\,GeV well.
To accommodate a hint of a structure in this region, fits are made by assuming a coherent sum of a perturbative QCD-driven energy power function and a Breit-Wigner (BW) function,
\begin{equation}\label{func11}
\sigma^{B}(\sqrt{s}) =\left|\sqrt{\frac{c_0\cdot\beta\cdot\cal{C}}{(\sqrt{s}-c_1)^{10}}} + e^{i\phi}BW(\sqrt{s})\sqrt{\frac{P(\sqrt{s})}{P(M)}}\right|^{2},
\end{equation}
where $P(\sqrt{s})$ is the two-body phase space factor, $\phi$ is the relative phase angle, taken as a free parameter, and 
\begin{equation}
BW(\sqrt{s}) =\frac{\sqrt{12\pi\Gamma_{ee}{\cal{B}}\Gamma}}{s-M^{2}+iM\Gamma},
\end{equation}
with mass $M$, width $\Gamma$, di-electron partial width of $\Gamma_{ee}$ and $\cal{B}$ is the branching fraction for the decay into the $\Xi^-\bar{\Xi}^+$ final state. This fit has $\chi^2/d.o.f$ = 0.03/2 and is given by the  red solid line in Fig.~\ref{bcsfit}. The statistical significance for this possible structure is estimated to be 2.4$\sigma$, when including the systematic uncertainties described above. The fitted mass and width are (2993 $\pm$ 28)\,MeV/c$^2$ and (88 $\pm$ 79)\,MeV, respectively, with $\phi =$ 2.4 $\pm$ 0.6 (4.3 $\pm$ 0.3) rad for constructive (destructive) interference condition~\cite{multisolution}, where the uncertainties are statistical only. The upper limit on the product $\Gamma_{ee}{\cal{B}}$ is estimated to be 0.1 (1.0) eV at the 90\% C.L. with mass and width of the structure fixed to the fitted value, using a Bayesian approach~\cite{ZHUYS} and taking into account the systematic uncertainties described above. The final function that is used to model the  $e^+e^-\to \Xi^-\bar{\Xi}^+$ process is inspired by the measured nucleon-pair production cross-section~\cite{iteration}, and therefore assumes a plateau near threshold. By taking into account the strong interaction near threshold instead of the Coulomb factor, the Born cross-section can be expressed as: 
\begin{equation}
\sigma^{B}(\sqrt{s}) =\frac{e^{a_0}\pi^2\alpha^3}{s\left[1-e^{-\frac{\pi\alpha_s}{\beta}}\right]\left[1+\left(\frac{\sqrt{s}-2m_{\Xi}}{a_1}\right)^{a_2}\right]},
\label{func2}
\end{equation}
where $a_0$, $a_1$, $a_2$ are fit parameters, $a_0$ is the normalization constant, $a_1$ is the QCD parameter near threshold, $a_2$ is the power-law related to the number of valence quarks and $\alpha_s$ is the running strong-coupling constant,
\begin{equation}
\alpha_s =\left[\frac{1}{\alpha_s(m_Z^2)}+\frac{7}{4\pi}\ln\left(\frac{s}{m_Z^2}\right)\right]^{-1}.
\end{equation}
Here $m_Z=91.1876$ GeV/c$^2$ is the mass of $Z$ boson and $\alpha_s(m_Z^2)=0.11856$ is the strong coupling constant at the $Z$ pole. The fit has $\chi^2/d.o.f = 1.15/5$ and is shown as the green dashed line in Fig.~\ref{bcsfit}. The inflection point of the plateau is around 3.0\,GeV, which is about 350\,MeV above threshold. The last two assumptions are seen to model the data better than the simple perturbative QCD-driven energy power function.

In summary, using a total integrated luminosity of 363.9\,pb$^{-1}$ $e^+e^-$ collision data collected with the BESIII detector at BEPCII, the Born cross-sections and effective form factors for the process $e^+e^-\to\Xi^-\bar{\Xi}^+$ are measured for the first time at c.m. energies between 2.644 and 3.08 GeV. A fit to the Born cross-section of $e^+e^-\to\Xi^-\bar{\Xi}^+$ is performed with several alternative assumptions: a QCD-driven energy power function, a variant of this that allows for a resonant structure at higher energies, and a variant that allows for a plateau near threshold.  All three approaches are compatible with the data, with the latter two providing somewhat better fits, although there is no significant evidence for either a resonance structure, or an unusual threshold behavior.  A resonant structure accommodated by the second assumption occurs at 3.0\,GeV with a  statistical significance of 2.4$\sigma$. The results of this analysis provide new and useful experimental information to understand the production mechanism for baryons with strangeness $S = -2$.

\label{sec:acknowledgement}
The BESIII collaboration thanks the staff of BEPCII and the IHEP computing center and the supercomputing center of USTC for their strong support. This work is supported in part by National Key Basic Research Program of China under Contract No. 2015CB856700; National Natural Science Foundation of China (NSFC) under Contracts Nos. 11335008, 11375170, 11475164, 11475169, 11605196, 11605198, 11625523, 11635010, 11705192, 11735014, 11822506, 11835012, 11905236, 11935015, 11935016, 11935018, 11950410506, 11961141012, 12075107, 12035013; the Chinese Academy of Sciences (CAS) Large-Scale Scientific Facility Program; Joint Large-Scale Scientific Facility Funds of the NSFC and CAS under Contracts Nos. U1532102, U1732263, U1832103, U1832207, U2032111; CAS Key Research Program of Frontier Sciences under Contracts Nos. QYZDJ-SSW-SLH003, QYZDJ-SSW-SLH040; 100 Talents Program of CAS; INPAC and Shanghai Key Laboratory for Particle Physics and Cosmology; ERC under Contract No. 758462; German Research Foundation DFG under Contracts Nos. 443159800, Collaborative Research Center CRC 1044, FOR 2359; Istituto Nazionale di Fisica Nucleare, Italy; Ministry of Development of Turkey under Contract No. DPT2006K-120470; National Science and Technology fund; Olle Engkvist Foundation under Contract No. 200-0605; STFC (United Kingdom); The Knut and Alice Wallenberg Foundation (Sweden) under Contract No. 2016.0157; The Royal Society, UK under Contracts Nos. DH140054, DH160214; The Swedish Research Council; U. S. Department of Energy under Contracts Nos. DE-FG02-05ER41374, DE-SC-0012069.

\end{document}